\documentclass[pra, showpacs, twocolumn, floatfix]{revtex4}
\usepackage{graphicx}
\usepackage{epstopdf}
\usepackage{amsmath, amsfonts, amssymb, bm}
\begin{document}
\title{Enhanced photon correlations due to strong laser-atom-cavity coupling}
\author{Viorel \surname{Ciornea}}
\author{Profirie \surname{Bardetski}}
\author{Mihai A. \surname{Macovei}}
\email{macovei@phys.asm.md}
\affiliation{Institute of Applied Physics, Academy of Sciences of Moldova,
Academiei str. 5, MD-2028 Chi\c{s}in\u{a}u, Moldova}
\date{\today}
\begin{abstract}
We investigate the resonant quantum dynamics of a laser-pumped real
or artificial two-level single-atom system embedded in a
leaking microcavity. We found that for stronger laser-atom-cavity couplings the generated
microcavity photons exhibit larger steady-state correlations. In
particular, the second- and third-order photon correlation functions
are greater than the corresponding ones obtained for an incoherent
light source, respectively. Furthermore, the emitted microcavity
photon flux is enhanced in comparison to weaker coupling cases.
\end{abstract}
\maketitle

\section{Introduction}
Correlation functions describing the coherence properties of
interacting or noninteracting quantum particles received
considerable attention 
\cite{glaub1,glaub2,sc_zb,luk,fl,fllk,mem,hart,mek,sant,carm,mjek,hicor1,hicor2,gk,ryd1,ryd2}.
These functions are used almost in all branches of modern physics.
Typically, standard sources of incoherent light generate photons
possessing a second-order correlation function $g^{(2)}(0)=2$
\cite{glaub1,glaub2,sc_zb}. Very recently, it was shown that photons
emitted by a single two-level system ultra-strong-coupled with a
thermal optical cavity show photon correlation functions $g^{(2)}(0)
>2$ or even smaller than unity exhibiting quantum light features,
i.e. sub-Poissonian photon statistics \cite{hart}. The effect occurs
when the atom-cavity coupling rate becomes comparable to the cavity 
resonance frequency. For such an interaction regime, the counter-rotating 
terms in the interaction Hamiltonian should be taken into account.
Therefore, various novel schemes generating highly correlated light 
are still important. Particularly, larger photon correlations are useful 
for a number of practical applications in many-body phenomena with 
strongly interacting photons \cite{luk} as well as in photonic quantum 
information processing \cite{fl,fllk,mem}. An ensemble of $N$ collectively 
interacting few-level atoms via an incoherent electromagnetic field 
reservoir generates light with $g^{(2)}(0)=4$ or even higher in 
the steady-state. However, the photon intensity is rather weak 
\cite{mek,sant}. Applying external coherent light sources, one can 
generate intenser photon fluxes proportional to $N$ or $N^{2}$ with, 
however, lower second-order photon correlations in the steady-state, 
i.e. $g^{(2)}(0) < 4$ \cite{carm,mjek}. Furthermore, higher-order photon 
correlations \cite{hicor1,hicor2} can be generated in a somehow more 
complicated setup involving nonlinear crystal superlattices \cite{gk}. 
Finally, correlated photon emission can be achieved from multiatom entangled 
Rydberg states, for instance \cite{ryd1,ryd2}. Thus, it becomes intriguing to 
find alternative ways to generate both an intense steady-state photon flux 
with enhanced photon-photon correlations.

Here, we demonstrate a scheme capable to generate a moderately
intense and highly correlated photon flux. In particular, the
obtained second- and third-order photon correlation functions in the
steady-state are several times larger than corresponding ones but
for a thermal light source.  The scheme is based on pumping a
two-level emitter embedded in a leaking optical  microcavity. At moderately
strong pumping, i.e. the respective Rabi frequency is larger than
the spontaneous and cavity decay rates, respectively, the
spontaneous scattered photons into free electromagnetic field modes
show the well-known Mollow spectrum \cite{moll} modified by the
cavity field \cite{gsa}. Now, if the laser-atom-cavity system is in
resonance with the central-band of the Mollow spectrum then the
side-bands will contribute to the atom-microcavity quantum dynamics
only via the non-secular terms. Under certain conditions, these
terms are responsible for enhancing the microcavity mean-photon
number as well as the second- and third-order photon correlations in
the steady-state. Furthermore, a moderate incoherent pumping of the
resonator mode does not modify the photon statistics considerably.
Notice the existence of an experiment measuring two- and
three-photon correlations in a strongly driven atom-cavity-system
for different parameters of interest \cite{rempe}. This makes our
results experimentally attractive.

The article is organized as follows. In Section II we describe the analytical approach
and the system of interest, while in Section III, we obtain the corresponding equations
of motion and describe the obtained results. The Summary is given in Section IV.

\section{Nonlinear quantum dynamics of a pumped qubit inside a microcavity}
The Hamiltonian describing a two-level real (or artificial) atomic system possessing the frequency $\omega_{0}$
and interacting with a coherent source of frequency $\omega_{L}$, and embedded in a microcavity of frequency $\omega_{c}$,
in a frame rotating at $\omega_{L}$, is:
\begin{eqnarray}
H = \hbar\Delta a^{\dagger}a + \hbar g(a^{\dagger}S^{-}+aS^{+}) +
 \hbar \Omega(S^{+} + S^{-}), \label{HM}
\end{eqnarray}
where we have assumed that $\omega_{0}=\omega_{L}$. In the Hamiltonian (\ref{HM})
the first term describes the cavity free energy with $\Delta=\omega_{c}-\omega_{L}$,
while the second one characterizes the interaction of the two-level emitter with the microcavity
mode via the coupling $g$. The third term considers the qubit's interaction with the laser field with
$\Omega$ being the corresponding Rabi frequency. The atomic bare-state operators $S^{+}=|2\rangle \langle 1|$ and
$S^{-}=[S^{+}]^{+}$ obey the commutation relations for su(2) algebra:
$[S^{+},S^{-}]=2S_{z}$ and $[S_{z},S^{\pm}]=\pm S^{\pm}$. Here, $S_{z}=(|2\rangle \langle 2| -
|1\rangle \langle 1|)/2$ is the bare-state inversion operator. $|2\rangle$ and $|1\rangle$ are the excited and
ground state of the qubit, respectively, while $a^{\dagger}$ and $a$ are the creation and the annihilation
operator of the electromagnetic field (EMF) in the resonator, and satisfy the standard bosonic commutation
relations, i.e., $[a,a^{\dagger}]=1$, and $[a,a]=[a^{\dagger},a^{\dagger}]=0$.

We are interested in the laser dominated regime where $\Omega \gg
\{\gamma,\kappa\}$ (here, $\gamma$ and $\kappa$ are the spontaneous
and cavity decay rates, respectively) and shall describe our system
using the dressed-states formalism \cite{sc_zb}: $ |1\rangle = (
|\bar 1\rangle + |\bar 2\rangle)/\sqrt{2}$, and $|2\rangle = ( |\bar
2\rangle  - |\bar 1\rangle)/\sqrt{2}$. Applying this transformation
in Eq.~(\ref{HM}) with $\Delta=0$, one arrive then at the following
dressed-state Hamiltonian in a frame rotating at the Rabi frequency
$\Omega$:
\begin{eqnarray}
H_{0} &=& \hbar g_{0}R_{z}(a + a^{\dagger}) + \hbar
g_{0}(R^{+}e^{2i\Omega t} - R^{-}e^{-2i\Omega t})\nonumber \\
&\times&(a^{\dagger} - a). \label{HI}
\end{eqnarray}
Here, $g_{0}=g/2$ while the new quasispin operators, i.e.
$R^{+}=|\bar 2\rangle\langle \bar 1|$, $R^{-}=[R^{+}]^{+}$ and
$R_{z}=|\bar 2\rangle\langle \bar 2| - |\bar 1\rangle\langle \bar
1|$ are operating in the dressed-state picture. They obey the
following commutation relations: $[R^{+},R^{-}]=R_{z}$ and
$[R_{z},R^{\pm}]=\pm 2R^{\pm}$. One can observe that the Hamiltonian
(\ref{HI}) can be separated into a time-independent part and a
time-dependent one containing fast oscillating terms. Therefore, the
time-dependent part, $H_{f}$, can be regarded as a perturbation to
the time-independent part when $\Omega > g_{0}$. One can apply then
the transformation: $\bar H=-\frac{i}{\hbar}H_{f}(t)\int dt H_{f}(t)$
\cite{gxl,guo,james} to arrive at the final time-independent Hamiltonian
characterizing the coherent evolution of the laser-atom-cavity system:
\begin{eqnarray}
H_{0} = \hbar g_{0}R_{z}(a^{\dagger} + a) + \hbar \beta
R_{z}a^{\dagger}a - \hbar\frac{\beta}{2}R_{z}(a^{\dagger 2} +
a^{2}), \label{H0}
\end{eqnarray}
where $\beta = g^{2}_{0}/\Omega$ and $\beta \ll g_{0}$. The
nonlinear terms proportional to $\beta$ in Eq.~(\ref{H0}) are due to
the non-secular contribution and are responsible for squeezed
one-atom lasing \cite{gxl} as well as non-classical EMF coherences
\cite{guo} in a different related setup.

In the Heisenberg picture, the master equation describing the
laser-dressed two-level qubit inside a leaking resonator and damped
via the vacuum modes of the surrounding EMF reservoir \cite{sc_zb}
is:
\begin{eqnarray}
&{}&\frac{d}{dt}\langle Q(t)\rangle  - \frac{i}{\hbar}\langle [H_{0},Q]\rangle
= -\Gamma_{0}\langle R_{z}[R_{z}, Q] \rangle \nonumber \\
&-& \Gamma\{ \langle R^{+}[R^{-},Q]\rangle + \langle R^{-}[R^{+},Q]\rangle \} \nonumber \\
&-& \kappa(1 + \bar n) \langle a^{\dagger}[a, Q]\rangle -\kappa \bar n \langle a[a^{\dagger}, Q]\rangle + H.c.. \label{ME}
\end{eqnarray}
Here, in general, for the non-Hermitian atomic or EMF operators $Q$,
the H.c. terms should be evaluated without conjugating $Q$, i.e. by
replacing $Q^{+}$ with $Q$ in the Hermitian conjugate parts.
Further, $\Gamma_{0} = \gamma/4$ and $\Gamma =
(\gamma+\gamma_{d})/4$ with $2\gamma$ being the single-atom
spontaneous decay rate, while $\gamma_{d}$ is the qubit dephasing
rate. Finally, $\bar n$ is the microcavity mean-photon number due to
incoherent pumping of the cavity mode. This is achieved via pumping
of the microcavity mode with a broadband laser field with its
spectral width larger than the resonator decay rate $\kappa$,
respectively. Note that in Eq.~(\ref{ME}), we have performed the
secular approximation in the spontaneous emission damping
\cite{mjek}.
\begin{figure}[t]
\includegraphics[width = 7cm]{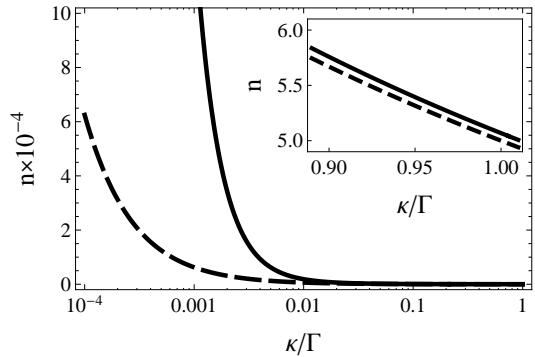}
\caption{\label{fig1} The steady-state dependences of the microcavity mean-photon number $n=\langle a^{\dagger}a \rangle_{s}$
versus the parameter $\kappa/\Gamma$. Dashed and solid curves are for $\beta/\Gamma = 0$ and $0.1$, respectively.
The inset shows the same behaviour for a particular range of $\kappa/\Gamma$. Here, $\bar n=0$ and $g_0/\Gamma=5$.}
\end{figure}

In the following section, we shall describe the microcavity photon statistics via the second- and
third-order photon correlation functions.
\section{Second- and third-order photon correlation functions}

The equations of motion for the variables of interest can be easily obtained from the Master Equation (\ref{ME}).
For instance, the steady-state value of the mean-photon number, i.e. $\langle a^{\dagger}a \rangle_s$, can be
extracted from the following system of linear steady-state equations:
\begin{eqnarray}
0&{=}&2 \kappa \langle a^{\dagger}a \rangle_s+i g_{0}\langle R_{z} a^{\dagger}\rangle_s-i g_{0}\langle R_{z} a\rangle_s\nonumber\\
&{-}&i \beta \langle R_{z} a^{\dagger 2}\rangle_s +i \beta \langle R_{z} a^{2}\rangle_s-2\kappa\bar n, \nonumber\\
0&{=}&(\kappa+4\Gamma) \langle R_{z} a^{\dagger} \rangle_s -i\beta\langle a^{\dagger}\rangle_s+i\beta\langle a\rangle_s-i g_{0}, \nonumber\\
0&{=}&(2\kappa+4\Gamma) \langle R_{z} a^{\dagger 2} \rangle_s - 2i g_{0}\langle a^{\dagger}\rangle_s-2i \beta \langle a^{\dagger 2}\rangle_s\nonumber\\
&{+}& i\beta(1+2\langle a^{\dagger}a \rangle_s),\nonumber\\
0&{=}&\kappa \langle a^{\dagger} \rangle_s - i \beta \langle R_{z} a^{\dagger}\rangle_s+i \beta \langle R_{z} a\rangle_s, \nonumber\\
0&{=}&2\kappa \langle a^{\dagger 2} \rangle_s - 2i g_{0}\langle R_{z} a^{\dagger}\rangle_s-2i \beta \langle R_{z} a^{\dagger 2}\rangle_s \nonumber\\
&{+}&2i\beta \langle R_{z} a^{\dagger}a \rangle_s, \nonumber\\
0&{=}&(2 \kappa+4\Gamma) \langle R_{z} a^{\dagger}a \rangle_s+i g_{0}\langle a^{\dagger}\rangle_s-i g_{0}\langle a\rangle_s\nonumber\\
&{-}& i\beta \langle  a^{\dagger 2}\rangle_s+i \beta \langle  a^{2}\rangle_s.\label{s1}
\end{eqnarray}
In the system of equations (\ref {s1}), and throughout the paper, we used the fact that the dressed-state inversion is zero in the steady-state,
i.e. $\langle R_{z}\rangle_{s}=0$, as well as the trivial condition $ R_z^2 =1$ which is the case for a single-atom system. By completing
the system of equations Eq.~(\ref {s1}) with the respective $H.c.$ equations, it takes a closed-form and can be exactly solved. In particular,
one of the solution of (\ref {s1}) represents the steady-state mean-photon number in the microcavity mode, namely:
\begin{eqnarray}
\langle a^{\dagger}a \rangle_{s} =\bar n+\frac{g_0^2}{\kappa (\kappa+4\Gamma)}+\frac{8g_0^2+\kappa(\kappa+4\Gamma)(1+2\bar n)}{2\kappa^2(\kappa+2\Gamma)(\kappa+4\Gamma)}\beta^2. \nonumber \\ \label{apm}
\end{eqnarray}
One can observe here, that the mean-photon number in the steady-state is enhanced due to both the incoherent pumping and the nonlinear
contribution proportional to $\beta^{2}$. This is clearly seen in Figure~(\ref{fig1}), where we plotted the microcavity mean-photon number,
$n=\langle a^{\dagger}a\rangle_{s}$, as a function of different relevant parameters. For $\kappa/\Gamma > 1$, the photon number in the
steady-state goes to $\bar n$ as the bad-cavity limit is achieved.
\begin{figure}[t]
\includegraphics[width = 7cm]{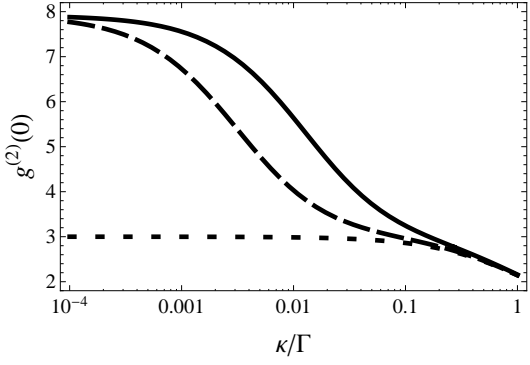}
\caption{\label{fig2} The steady-state dependences of the microcavity second-order photon correlation function $g^{(2)}(0)$ versus
the parameter $\kappa/\Gamma$. Short-dashed, long-dashed and solid curves are for $\beta/\Gamma = 0$,
$0.05$, and $0.1$, respectively. Other parameters are: $\bar n=0$ and $g_0/\Gamma=5$.}
\end{figure}
\begin{figure}[b]
\includegraphics[width = 7cm]{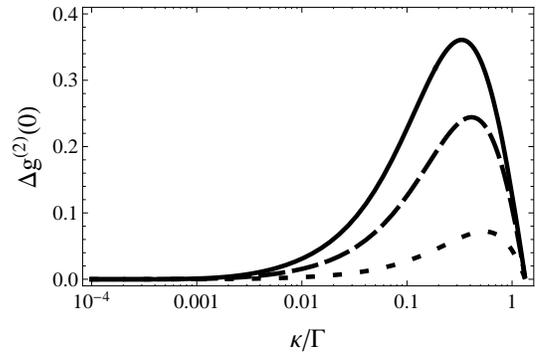}
\caption{\label{fig3} The variance of the second-order correlation functions
$\Delta g^{(2)}(0)=g^{(2)}(0)|_{\bar n\neq 0}-g^{(2)}(0)|_{\bar n=0}$ as a function of $\kappa/\Gamma$.
Short-dashed, long-dashed and solid lines are for $\bar n=1$, $5$, and $10$, respectively.
Here, $g_0/\Gamma=5$ and $\beta/\Gamma = 0.05$.}
\end{figure}

Further, we focus on the degree of second- and third-order coherences of microcavity photons defined, respectively, as \cite{glaub1,glaub2}:
\begin{equation}
g^{(2)}(0)=\frac{\langle a^{\dagger 2}a^2 \rangle_{s}}{\langle a^{\dagger}a \rangle_{s}^2}, ~~{\rm and}~~
g^{(3)}(0)=\frac{\langle a^{\dagger 3}a^3 \rangle_{s}}{\langle a^{\dagger}a \rangle_{s}^3}. \label{crr}
\end{equation}
The unnormalized k-order correlation function $\langle a^{\dagger k}a^{k}\rangle_{s}$ describes the probability of a k-photon detection, simultaneously.
To obtain the steady-state expressions for correlation functions given in Eq.~(\ref{crr}), the system of equations (\ref{s1}) have to be completed with
additional equations obtained with the help of Eq.~(\ref{ME}). In particular, the unnormalized steady-state second-order photon correlation function can
be obtained from the following system of equations (and the corresponding $H.c.$ expressions):
\begin{eqnarray}
0&{=}&4 \kappa \langle a^{\dagger 2}a^2 \rangle_s+2i g_{0}\langle R_{z} a^{\dagger 2}a\rangle_s-2i g_{0}\langle R_{z} a^{\dagger}a^2\rangle_s\nonumber\\
&{-}&i \beta \big(\langle R_{z} a^{\dagger 2}\rangle_s+2\langle R_{z} a^{\dagger 3}a\rangle_s\nonumber\\
&{-}& \langle R_{z} a^{2}\rangle_s-2\langle R_{z} a^{\dagger}a^3\rangle_s\big)-8\kappa\bar n\langle a^{\dagger}a\rangle_s, \nonumber\\
0&{=}&(3\kappa+4\Gamma)\langle R_za^{\dagger 2}a\rangle_s+i g_{0}\langle a^{\dagger 2}\rangle_s-2i g_{0}\langle a^{\dagger}a\rangle_s\nonumber\\
&{-}&i \beta \big(\langle a^{\dagger 2}a\rangle_s+\langle a^{\dagger 3}\rangle_s-\langle a\rangle_s-2\langle a^{\dagger}a^2\rangle_s\big)\nonumber\\
&{-}&4\kappa\bar n\langle R_za^{\dagger}\rangle_s,\nonumber\\
0&{=}&(4\kappa+4\Gamma)\langle R_za^{\dagger 3}a\rangle_s+i g_{0}\langle a^{\dagger 3}\rangle_s-3i g_{0}\langle a^{\dagger 2}a\rangle_s\nonumber\\
&{-}& i\beta \big(2\langle a^{\dagger 3}a\rangle_s+\langle a^{\dagger 4}\rangle_s-3\langle a^{\dagger 2}a^2\rangle_s-3\langle a^{\dagger}a\rangle_s\big)\nonumber\\
&{-}&6\kappa\bar n\langle R_za^{\dagger 2}\rangle_s,\nonumber\\
0&{=}&3\kappa\langle a^{\dagger}a^2\rangle_s+2 ig_{0}\langle R_za^{\dagger}a\rangle_s-ig_{0}\langle R_za^2\rangle_s\nonumber\\
&{+}& i\beta \big(\langle R_za^{\dagger}a^2\rangle_s-\langle R_za^{\dagger}\rangle_s\nonumber\\
&{-}&2\langle R_za^{\dagger 2}a\rangle_s+\langle R_za^3\rangle_s\big)-4\kappa\bar n\langle a\rangle_s,\nonumber\\
0&{=}&3\kappa\langle a^{\dagger 3}\rangle_s-3ig_{0}\langle R_za^{\dagger 2}\rangle_s\nonumber\\
&{+}&3i\beta \big(\langle R_za^{\dagger 2}a\rangle_s+\langle R_za^{\dagger}\rangle_s-\langle R_za^{\dagger 3}\rangle_s\big),\nonumber\\
0&{=}&4\kappa\langle a^{\dagger 4}\rangle_s-4ig_{0}\langle R_za^{\dagger 3}\rangle_s\nonumber\\
&{+}&2i\beta\big(2\langle R_za^{\dagger 3}a\rangle_s+3 \langle R_za^{\dagger 2}\rangle_s-2\langle R_za^{\dagger 4}\rangle_s\big),\nonumber\\
0&{=}&4\kappa\langle a^{\dagger 3}a\rangle_s+ig_{0}\langle R_za^{\dagger 3}\rangle_s-3ig_{0}\langle R_za^{\dagger 2}a\rangle_s\nonumber\\
&{-}&i \beta \big(2\langle R_za^{\dagger 3}a\rangle_s+\langle R_za^{\dagger 4}\rangle_s\nonumber\\
&{-}&3\langle R_za^{\dagger 2}a^2\rangle_s-3\langle R_za^{\dagger}a\rangle_s\big)-6\kappa\bar n\langle a^{\dagger 2}\rangle_s,\nonumber\\
0&{=}&(3\kappa+4\Gamma)\langle R_za^{\dagger 3}\rangle_s-3ig_{0}\langle a^{\dagger 2}\rangle_s\nonumber\\
&{+}&3i\beta \big(\langle a^{\dagger 2}a\rangle_s+\langle a^{\dagger}\rangle_s-\langle a^{\dagger 3}\rangle_s\big),\nonumber\\
0&{=}&(4\kappa+4\Gamma)\langle R_za^{\dagger 4}\rangle_s-4ig_{0}\langle a^{\dagger 3}\rangle_s\nonumber\\
&{+}&2i\beta\big(2\langle a^{\dagger 3}a\rangle_s+3 \langle a^{\dagger 2}\rangle_s-2\langle a^{\dagger 4}\rangle_s\big),\nonumber\\
0&{=}&(4 \kappa+4\Gamma) \langle R_z a^{\dagger 2}a^2 \rangle_s+2i g_{0}\langle a^{\dagger 2}a\rangle_s-2i g_{0}\langle a^{\dagger}a^2\rangle_s\nonumber\\
&{-}&i \beta \big(\langle a^{\dagger 2}\rangle_s+2\langle a^{\dagger 3}a\rangle_s\nonumber\\
&{-}& \langle a^{2}\rangle_s-2\langle a^{\dagger}a^3\rangle_s\big)-8\kappa\bar n\langle R_za^{\dagger}a\rangle_s. \label{s2}
\end{eqnarray}
The equation of motions required to obtain the third-order correlation function are given in the Appendix B.

Taking into account the systems of equations (\ref {s1}) and (\ref {s2}), one can obtain the steady-state
expression for the second-order coherence function, that is:
\begin{equation}
g^{(2)}(0)=\frac{(\kappa + 2 \Gamma) (\kappa + 4 \Gamma)}{3 (\kappa+\Gamma )^2 (3\kappa+4 \Gamma)^2}
\frac{ A+B\beta^2+C\beta^4}{\tilde A+\tilde B\beta^2+\tilde C\beta^4}, \label{g2}
\end{equation}
where
\begin{figure}[t]
\includegraphics[width = 7cm]{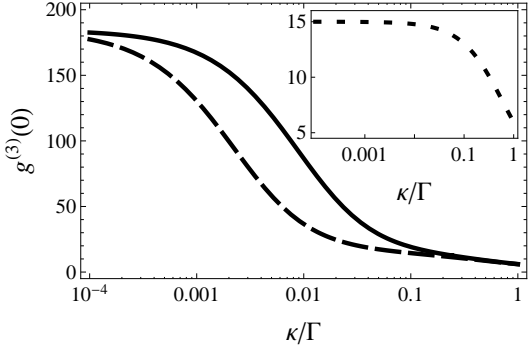}
\caption{\label{fig4} The steady-state dependences of the microcavity third-order photon coherence function $g^{(3)}(0)$ as a function of $\kappa/\Gamma$.
Solid and long-dashed curves are for $\beta/\Gamma = 0.1$ and $0.05$, respectively. The inset shows the case of $\beta/\Gamma = 0$.
Other parameters are: $\bar n=0$ and $g_0/\Gamma=5$.}
\end{figure}
\begin{eqnarray}
A&{=}&A_0+A_1\bar n+A_2\bar n^2, ~ \tilde A=\tilde A_0+\tilde A_1\bar n+\tilde A_2\bar n^2,\nonumber\\
B&{=}&B_0+B_1\bar n+B_2\bar n^2, ~ \tilde B=\tilde B_0+\tilde B_1\bar n+\tilde B_2\bar n^2,\nonumber\\
C&{=}&C_0+C_1\bar n+C_2\bar n^2, ~ \tilde C=\tilde C_0+\tilde C_1\bar n+\tilde C_2\bar n^2, \nonumber
\end{eqnarray}
while the missing parameters are given in Eqs.~(\ref{a1}), in the Appendix A.  Here, again, one can observe that the nonlinear term $\beta$
modifies the second-order photon correlation. In particular, when $\beta=0$ one obtains:
\begin{eqnarray}
g^{(2)}(0)=2 + \frac{g^{4}_{0}(4\Gamma - 3\kappa)}{(4\Gamma + 3\kappa)\big(g^{2}_{0} + \kappa(\kappa + 4\Gamma)\bar n \big)^2}, \label{g2b}
\end{eqnarray}
while for $\{g_{0},\beta\}=0$ one has $g^{(2)}(0)=2$, that is, we recover the incoherent-source result for a second-order correlation
function. For smaller values of $\kappa/\Gamma$ and $\beta \not= 0$, the second-order coherence function tends to a constant value:
\begin{equation}
\lim_{\kappa \to 0}g^{(2)}(0)=\frac{95}{12} \approx 7.91,
\end{equation}
while for $\beta=0$ and $g_{0} \not =0$ we have:
\begin{equation}
\lim_{\kappa \to 0}g^{(2)}(0)=3.
\end{equation}
These behaviors are shown in Fig.~(\ref{fig2}), plotted with the help of Eq.~(\ref{g2}). Thus, for particular values of involved parameters one can
obtain larger photon correlations, i.e. $g^{(2)}(0) \gg 2$, as well as bigger photon numbers (see Fig.~\ref{fig2} and Fig.~\ref{fig1}, respectively)
due to nonlinear terms proportional to $\beta$ in the interaction Hamiltonian (\ref{H0}). 
To elucidate the role played by the incoherent pumping, in Fig.~(\ref{fig3}), we depict the difference
$\Delta g^{(2)}(0)=g^{(2)}(0)|_{\bar n\neq 0}-g^{(2)}(0)|_{\bar n=0}$ as a function of $\kappa/\Gamma$. For very small values of
$\kappa/\Gamma$, a moderate incoherent pumping almost does not affect the photon coherences, while when the ratio $\kappa/\Gamma$
increases the photon statistics modifies accordingly (see Fig.~\ref{fig3} and Eqs.~\ref{g2} and \ref{g2b}).
\begin{figure}[t]
\includegraphics[width = 7cm]{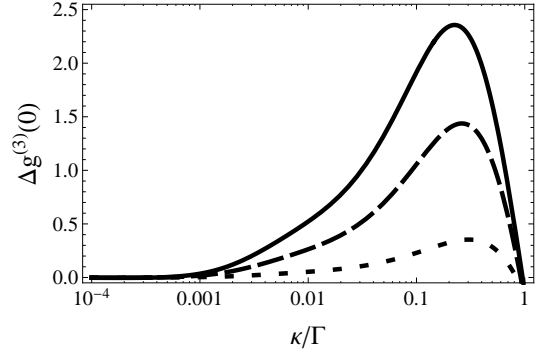}
\caption{\label{fig5} The same as in Fig.~(\ref{fig3}), but for $\Delta g^{(3)}(0)=g^{(3)}(0)|_{\bar n\neq 0}-g^{(3)}(0)|_{\bar n=0}$.}
\end{figure}

In order to understand that the photon correlations indeed are enhanced due to the nonlinearity characterized by $\beta$, in Fig.~(\ref{fig4}), we plot
the third-order microcavity photon correlation function $g^{(3)}(0)$ (with the help of the corresponding equations, i.e. Eqs.~(\ref{b1}), given in the Appendix B).
Remarkable, the third-order photon correlations are enhanced as well due to presence of $\beta$. In particular, for an incoherent bath,
i.e. $g_{0}=0$, $g^{(3)}(0)=6$ while for $\beta/\Gamma=0$ the third-order steady-state photon coherence function is given by the expression:
\begin{eqnarray}
g^{(3)}(0)&=& 6 + \frac{12g^{6}_{0} \kappa (5 \kappa - 12 \Gamma) (5 \kappa + 4 \Gamma)^{-1}}{(3 \kappa + 4 \Gamma) (g^{2}_{0} +
   \kappa (\kappa + 4 \Gamma) \bar n)^3}\nonumber\\
&{+}&\frac{9g^{4}_{0} (-3 \kappa + 4 \Gamma)}{(3 \kappa + 4 \Gamma) (g^{2}_{0} + \kappa (\kappa + 4 \Gamma) \bar n)^2}. \label{g3}
\end{eqnarray}
In general, i.e. for $\beta \not=0$, the expression for $g^{(3)}(0)$ is too complicated and it is not shown here.
However, for smaller values of $\kappa/\Gamma$ and $\beta \not=0$ one has:
\begin{equation}
\lim_{\kappa \to 0}g^{(3)}(0)=\frac{33203}{180} \approx 184.46. \label{l1}
\end{equation}
If $\beta=0$ and $g_{0} \not =0$ we have:
\begin{equation}
\lim_{\kappa \to 0}g^{(3)}(0)=15. \label{l2}
\end{equation}
Thus, in both limits described by expressions (\ref{l1},\ref{l2}), the third-order correlation function is bigger than
the corresponding one but for an incoherent-type light source, i.e. $g^{(3)}(0) >  6$ (see, also, Fig.~\ref{fig4}).
The influence of a moderate incoherent pumping into the cavity mode on third-order photon statistics is shown in
Fig.~(\ref{fig5}). Similar to Fig.~(\ref{fig3}), very small values of $\kappa/\Gamma$ do not modify the third-order
photon correlations in presence of incoherent photons $\bar n$. $g^{(3)}(0)$ slightly changes with increasing
$\bar n$ and $\kappa/\Gamma$. Stronger incoherent pumping will lead to a photon statistics which is typical for an
incoherent-light source. Furthermore, the photon statistics does not depend on $g_{0}$ for $\{\bar n, \beta\} = 0$
(see, for instance, Eq.~\ref{g2b} and Eq.~\ref{g3}). Finally, one can conjecture that even higher-order photon correlation
functions will behave similarly to those shown in Fig.~(\ref{fig2}) - Fig.~(\ref{fig5}) with, however, larger magnitudes.

\section{Summary}
In summary, we have investigated the correlations functions of microcavity photons generated due to coherent interaction of a two-level
qubit with an external laser field as well as a microcavity mode and in the presence of both the spontaneous emission and cavity damping,
respectively. In addition, the cavity mode is pumped incoherently. In the intense-field limit, i.e. the involved Rabi frequency is larger than
the spontaneous and cavity decay rates, respectively, we found enhanced second- and third-order photon correlations. These correlation
functions  are larger than the corresponding ones but for an incoherent-light source. The photon correlation enhancement is due to the
non-secular contribution in the coherent dressed-state Hamiltonian describing the pumped atom interacting with the microcavity mode. This
contribution is relevant for stronger laser-atom-cavity interactions.
\begin{acknowledgements}
We are grateful to financial support via the research grant Nr. 13.820.05.07/GF.
\end{acknowledgements}

\appendix
\section{The parameters entering in the second-order photon correlation function}
Here are the parameters entering in Eq.~(\ref{g2}).

\begin{eqnarray}
A_0&{=}&36 g^{4}_{0} \kappa^2 (\Gamma + \kappa)^2 (2 \Gamma
+ \kappa) (4 \Gamma + 3 \kappa),\nonumber\\
A_1&{=}&48 g^{2}_{0} \kappa^3 (\Gamma + \kappa)^2 (2 \Gamma
+ \kappa) (4 \Gamma + 3 \kappa)^2,\nonumber\\
A_2&{=}&24 \kappa^4 (\Gamma + \kappa)^2 (2 \Gamma +
\kappa) (4 \Gamma + \kappa) (4 \Gamma +
   3 \kappa)^2,\nonumber\\
\tilde A_0&{=}&4 g^{4}_{0} \kappa^2 (2 \Gamma + \kappa)^2,\nonumber\\
\tilde A_1&{=}&8 g^{2}_{0} \kappa^3 (2 \Gamma + \kappa)^2 (4 \Gamma
+ \kappa),\nonumber\\
\tilde A_2&{=}&4 \kappa^4 (2 \Gamma + \kappa)^2 (4 \Gamma +
\kappa)^2,\nonumber\\
B_{0}&{=}&\kappa (\Gamma + \kappa)\big (3 \kappa^2 (\Gamma
+ \kappa) (4 \Gamma + \kappa) (4 \Gamma +
      3 \kappa)^2 \nonumber\\
&{+}& 32 g^{4}_{0} (50 \Gamma^2 + 74 \Gamma \kappa + 27 \kappa^2) \nonumber\\
&{+}&  4g^{2}_{0} \kappa (4 \Gamma + 3 \kappa) (100 \Gamma^2 + 139 \Gamma \kappa + 45 \kappa^2) \big),\nonumber\\
B_{1}&{=}&4 \kappa^2 (\Gamma + \kappa) (4 \Gamma + 3 \kappa) \big(g^{2}_{0}(392 \Gamma^2 + 614 \Gamma \kappa  \nonumber\\
&{+}& 234 \kappa^2)+ 9 \kappa (\Gamma + \kappa) (4 \Gamma + \kappa) (4 \Gamma + 3 \kappa) \big),\nonumber
\end{eqnarray}
\begin{eqnarray}
B_2&{=}&60 \kappa^3 (\Gamma + \kappa)^2 (4 \Gamma + \kappa) (4 \Gamma + 3 \kappa)^2,\nonumber\\
\tilde B_0&{=}&4 \kappa g^{2}_{0}(2 \Gamma + \kappa) (8 g^{2}_{0} +  \kappa (4 \Gamma + \kappa)),\nonumber\\
\tilde B_1&{=}&4 \kappa^2 (2 \Gamma + \kappa) (4 \Gamma + \kappa) (10 g^{2}_{0} + \kappa (4 \Gamma + \kappa)), \nonumber\\
\tilde B_2&{=}&8 \kappa^3 (2 \Gamma + \kappa) (4 \Gamma + \kappa)^2,\nonumber\\
C_0&{=}&9 \kappa^2 (\Gamma + \kappa) (4 \Gamma + \kappa) (4 \Gamma + 3 \kappa)^2 \nonumber\\
&{+}&  16 g^{4}_{0} (190 \Gamma^2 + 289 \Gamma \kappa + 108 \kappa^2) \nonumber\\
&{+}&  4 g^{2}_{0} \kappa (4 \Gamma + 3 \kappa) (220 \Gamma^2 + 319 \Gamma \kappa + 108 \kappa^2), \nonumber \\
C_{1}&{=}&4 \kappa (4 \Gamma + 3 \kappa)\big (9 \kappa (\Gamma + \kappa) (4 \Gamma + \kappa) (4 \Gamma + 3 \kappa) \nonumber\\
&{+}&  g^{2}_{0} (440 \Gamma^2 + 638 \Gamma \kappa + 216 \kappa^2) \big), \nonumber\\
C_2&{=}&36 \kappa^2 (\Gamma + \kappa) (4 \Gamma + \kappa) (4 \Gamma + 3 \kappa)^2, \nonumber \\
\tilde C_0&{=}&(8 g^{2}_{0} + \kappa (4 \Gamma + \kappa))^2,\nonumber\\
\tilde C_1&{=}&4 \kappa (4 \Gamma + \kappa) (8g^{2}_{0} + \kappa (4 \Gamma + \kappa)), \nonumber\\
\tilde C_2&{=}&4 \kappa^2 (4 \Gamma + \kappa)^2. \label{a1}
\end{eqnarray}

\section{Equations of motion for the third-order photon coherences}
To obtain the third-order correlation function $g^{(3)}(0)$ the system of linear equations (\ref {s1}) \& (\ref {s2}) should be completed
with the following equations of motion (and the corresponding $H.c.$ equations):
\begin{eqnarray}
0&{=}&6 \kappa \langle a^{\dagger 3}a^3 \rangle_s+3i g_{0}\langle R_{z} a^{\dagger 3}a^2\rangle_s-3i g_{0}\langle R_{z} a^{\dagger 2}a^3\rangle_s\nonumber\\
&{}&-3i \beta \big(\langle R_{z} a^{\dagger 4}a^2\rangle_s+\langle R_{z} a^{\dagger 3}a\rangle_s\nonumber\\
&{}&- \langle R_{z} a^{\dagger 2}a^4\rangle_s - \langle R_{z} a^{\dagger}a^3\rangle_s\big) -18\kappa \bar n \langle a^{\dagger 2}a^2 \rangle_s, \nonumber \\
0&{=}&(6\kappa+4\Gamma)\langle R_za^{\dagger 3}a^3 \rangle_s+3i g_{0}\langle a^{\dagger 3}a^2\rangle_s\nonumber\\
&{}&-3i g_{0}\langle a^{\dagger 2}a^3\rangle_s -3 i \beta \big(\langle a^{\dagger 4}a^2\rangle_s+\langle  a^{\dagger 3}a\rangle_s\nonumber\\
&{}&- \langle  a^{\dagger 2}a^4\rangle_s - \langle a^{\dagger}a^3\rangle_s\big)-18\kappa \bar n \langle R_z a^{\dagger 2}a^2 \rangle_s, \nonumber \\
0&{=}&(5\kappa+4\Gamma)\langle R_za^{\dagger 3}a^2\rangle_s+2ig_{0}\langle a^{\dagger 3}a\rangle_s-3i g_{0}\langle a^{\dagger 2}a^2\rangle_s\nonumber\\
&{}&-i \beta \big(\langle a^{\dagger 3}a^2\rangle_s+\langle a^{\dagger 3}\rangle_s+2\langle a^{\dagger 4}a\rangle_s\nonumber\\
&{}&-3\langle a^{\dagger 2}a^3\rangle_s-3\langle a^{\dagger}a^2\rangle_s\big)-12\kappa \bar n \langle R_z a^{\dagger 2}a \rangle_s,\nonumber\\
0&{=}&5\kappa\langle a^{\dagger 3}a^2\rangle_s+2ig_{0}\langle R_za^{\dagger 3}a\rangle_s-3i g_{0}\langle R_za^{\dagger 2}a^2\rangle_s\nonumber\\
&{}&-i \beta \big(\langle R_za^{\dagger 3}a^2\rangle_s+\langle R_za^{\dagger 3}\rangle_s+2\langle R_za^{\dagger 4}a\rangle_s\nonumber\\
&{}&-3\langle R_za^{\dagger 2}a^3\rangle_s-3\langle R_za^{\dagger}a^2\rangle_s\big)-12\kappa \bar n \langle a^{\dagger 2}a \rangle_s,\nonumber \\
0&{=}&(6\kappa+4\Gamma)\langle R_za^{\dagger 4}a^2\rangle_s+2ig_{0}\langle a^{\dagger 4}a\rangle_s-4ig_{0}\langle a^{\dagger 3}a^2\rangle_s\nonumber\\
&{}&-i \beta \big(2\langle a^{\dagger 4}a^2\rangle_s+\langle a^{\dagger 4}\rangle_s+2\langle a^{\dagger 5}a\rangle_s\nonumber\\
&{}&-4\langle a^{\dagger 3}a^3\rangle_s-6\langle a^{\dagger 2}a^2\rangle_s\big)-16\kappa \bar n \langle R_z a^{\dagger 3}a \rangle_s,\nonumber \\
0&{=}&6\kappa\langle a^{\dagger 4}a^2\rangle_s+2ig_{0}\langle R_za^{\dagger 4}a\rangle_s-4ig_{0}\langle R_za^{\dagger 3}a^2\rangle_s\nonumber\\
&{}&-i \beta \big(2\langle R_za^{\dagger 4}a^2\rangle_s+\langle R_za^{\dagger 4}\rangle_s+2\langle R_za^{\dagger 5}a\rangle_s\nonumber\\
&{}&-4\langle R_za^{\dagger 3}a^3\rangle_s-6\langle R_za^{\dagger 2}a^2\rangle_s\big)-16\kappa \bar n \langle a^{\dagger 3}a \rangle_s,\nonumber  \\
0&{=}&(5\kappa+4\Gamma)\langle R_za^{\dagger 4}a\rangle_s+ig_{0}\langle a^{\dagger 4}\rangle_s-4ig_{0}\langle a^{\dagger 3}a\rangle_s\nonumber\\
&{}&-i \beta \big(3\langle a^{\dagger 4}a\rangle_s+\langle a^{\dagger 5}\rangle_s
-4\langle a^{\dagger 3}a^2\rangle_s-6\langle a^{\dagger 2}a\rangle_s\big),\nonumber\\
&{}&-8\kappa \bar n \langle R_z a^{\dagger 3} \rangle_s,\nonumber 
\end{eqnarray}
\begin{eqnarray}
0&{=}&5\kappa\langle a^{\dagger 4}a\rangle_s+ig_{0}\langle R_za^{\dagger 4}\rangle_s-4ig_{0}\langle R_za^{\dagger 3}a\rangle_s\nonumber\\
&{}&-i \beta \big(3\langle R_za^{\dagger 4}a\rangle_s+\langle R_za^{\dagger 5}\rangle_s\nonumber\\
&{}&-4\langle R_za^{\dagger 3}a^2\rangle_s-6\langle R_za^{\dagger 2}a\rangle_s\big) -8\kappa \bar n \langle a^{\dagger 3} \rangle_s,\nonumber \\
0&{=}&(6\kappa+4\Gamma)\langle R_za^{\dagger 5}a\rangle_s+ig_{0}\langle a^{\dagger 5}\rangle_s-5ig_{0}\langle a^{\dagger 4}a\rangle_s\nonumber\\
&{}&-i \beta \big(4\langle a^{\dagger 5}a\rangle_s+\langle a^{\dagger 6}\rangle_s
-5\langle a^{\dagger 4}a^2\rangle_s-10\langle a^{\dagger 3}a\rangle_s\big)\nonumber \\
&{}&-10\kappa \bar n \langle R_z a^{\dagger 4} \rangle_s,\nonumber\\
0&{=}&6\kappa\langle a^{\dagger 5}a\rangle_s+ig_{0}\langle R_za^{\dagger 5}\rangle_s-5ig_{0}\langle R_za^{\dagger 4}a\rangle_s\nonumber\\
&{}&-i \beta \big(4\langle R_za^{\dagger 5}a\rangle_s+\langle R_za^{\dagger 6}\rangle_s\nonumber\\
&{}&-5\langle R_za^{\dagger 4}a^2\rangle_s-10\langle R_za^{\dagger 3}a\rangle_s\big)-10\kappa \bar n \langle  a^{\dagger 4} \rangle_s,\nonumber \\
0&{=}&(5\kappa+4\Gamma)\langle R_za^{\dagger 5}\rangle_s-5ig_{0}\langle a^{\dagger 4}\rangle_s\nonumber\\
&{}&-i \beta \big(5\langle a^{\dagger 5}\rangle_s-5\langle a^{\dagger 4}a\rangle_s-10\langle a^{\dagger 3}\rangle_s\big),\nonumber \\
0&{=}&5\kappa\langle a^{\dagger 5}\rangle_s-5ig_{0}\langle R_za^{\dagger 4}\rangle_s\nonumber\\
&{}&-i \beta \big(5\langle R_za^{\dagger 5}\rangle_s-5\langle R_za^{\dagger 4}a\rangle_s-10\langle R_za^{\dagger 3}\rangle_s\big),\nonumber\\
0&{=}&(6\kappa+4\Gamma)\langle R_za^{\dagger 6}\rangle_s-6ig_{0}\langle a^{\dagger 5}\rangle_s\nonumber\\
&{}&-i \beta \big(6\langle a^{\dagger 6}\rangle_s-6\langle a^{\dagger 5}a\rangle_s-15\langle a^{\dagger 4}\rangle_s\big),\nonumber\\
0&{=}&6\kappa\langle a^{\dagger 6}\rangle_s-6ig_{0}\langle R_za^{\dagger 5}\rangle_s\nonumber\\
&{}&-i \beta \big(6\langle R_za^{\dagger 6}\rangle_s-6\langle R_za^{\dagger 5}a\rangle_s-15\langle R_za^{\dagger 4}\rangle_s\big). \nonumber \\
\label{b1}
\end{eqnarray}



\begin{thebibliography}{33}
\bibitem{glaub1} R. J. Glauber, Phys. Rev. {\bf 130}, 2529 (1963).

\bibitem{glaub2} R. J. Glauber, Rev. Mod. Phys. {\bf 78}, 1267 (2006).

\bibitem{sc_zb} M. O. Scully and M. S. Zubairy, {\it Quantum Optics} (Cambridge
University Press, Cambridge, 1997).

\bibitem{hart} A. Ridolfo, S. Savasta, M. Hartmann, Phys. Rev. Lett. {\bf 110}, 163601 (2013).

\bibitem{luk} D. E. Chang, V. Gritsev, G. Morigi, V. Vuletic,  M. D. Lukin,  E. A. Demler, 
Nature Phys. {\bf 4}, 884 (2008).

\bibitem{fl} I. Friedler, D. Petrosyan, M. Fleischhauer, G. Kurizki,
Phys. Rev. A {\bf 72}, 043803 (2005). 

\bibitem{fllk} A. V. Gorshkov, J. Otterbach, M. Fleischhauer, Th. Pohl, M. D. Lukin,
Phys. Rev. Lett. {\bf 107}, 133602 (2011). 

\bibitem{mem} A. I. Lvovsky, B. C. Sanders,  W. Tittel,
Nature Photon. {\bf 3}, 706 (2009).

\bibitem{mek} M. Macovei, J. Evers, C. H. Keitel, Phys. Rev. A {\bf 72}, 063809 (2005).

\bibitem{sant} A. Auffeves, D. Gerace, S. Portolan, A. Drezet, M. Santos,
 New Jr. of  Phys. {\bf 13}, 093020  (2011).

\bibitem{carm}  D. Norris, A. Cimmarusti, L. Orozco, P. Barberis-Blostein, H. Carmichael,
Phys. Rev. A {\bf 86}, 053816 (2012).

\bibitem{mjek} M. Macovei, C. H. Keitel, Phys. Rev. B {\bf 75}, 245325 (2007);
L. Jin, J. Evers, M. Macovei, Phys. Rev. A {\bf 84}, 043812 (2011).

\bibitem{hicor1} A. Allevi, S. Olivares, M. Bondani, Phys. Rev. A {\bf 85}, 063835 (2012).

\bibitem{hicor2} E. del Valle, A. Gonzalez-Tudela, F. Laussy, C. Tejedor, M. Hartmann, Phys. Rev. Lett. {\bf 109}, 183601 (2012).

\bibitem{gk} D. A. Antonosyan, T. V. Gevorgyan, G. Yu. Kryuchkyan, Phys. Rev. A {\bf 83}, 043807 (2011).

\bibitem{ryd1} J. Pritchard, C. Adams, K. Molmer, Phys. Rev. Lett. {\bf 108}, 043601 (2012).

\bibitem{ryd2} F. Bariani, Y. Dudin, T. Kennedy, A. Kuzmich, Phys. Rev. Lett. {\bf 108}, 030501 (2012).

\bibitem{moll} B. R. Mollow, Phys. Rev. {\bf 188}, 1969 (1969).

\bibitem{gsa} G. S. Agarwal, Phys. Rev. A {\bf 41}, 2886 (1990).

\bibitem{rempe} M. Koch, C. Sames, M. Balbach, H. Chibani, A. Kubanek, K. Murr, T. Wilk, G. Rempe, Phys. Rev. Lett. {\bf 107}, 023601 (2011).

\bibitem{gxl} R. Tan, G.-x. Li, Z. Ficek, Phys. Rev. A {\bf 78}, 023833 (2008).

\bibitem{guo} X. Guo, Sh. L\"{u}, Zh. Ren, Jr. of Phys. B: At., Mol. Opt. Phys. {\bf 43}, 225401 (2010).

\bibitem{james} D. F. V. James, Fort. Phys. {\bf 48}, 823 (2000).
\end{thebibliography}
\end{document}